\documentclass[twocolumn,showpacs,preprintnumbers,amsmath,amssymb,nofootinbib]{revtex4}
\usepackage{dcolumn}
\usepackage{bm}
\usepackage{tabularx}
\usepackage{array}
\usepackage{graphics}
\usepackage{graphicx}
\usepackage{epsfig}
\usepackage{amsmath}
\usepackage{amssymb}
\usepackage{citesort}
\usepackage{verbatim}

\newcommand{\msbar}{{\overline{\rm MS}}}
\newcommand{\ms}{{m_{\rm sea}}}
\newcommand{\mv}{{m_{\rm val.}}}
\newcommand{\rs}{{r_{\rm sea}}}
\newcommand{\rv}{{r_{\rm val.}}}
\newcommand{\ns}{{N_{\rm sea}}}

\newcommand{\bea}{\begin{eqnarray}}
\newcommand{\eea}{\end{eqnarray}}
\newcommand{\beq}{\begin{equation}}
\newcommand{\eeq}{\end{equation}}
\newcommand{\gev}{{\rm GeV}}

\newcommand{\pdir}{p\kern -5.2pt\raise 0.2ex\hbox {/}}
\newcommand{\vdir}{v\kern -5.75pt\raise 0.15ex\hbox {/}}
\newcommand{\kdir}{k\kern -5.75pt\raise 0.15ex\hbox {/}}
\newcommand{\epsdir}{\epsilon\kern -5.0pt\raise 0.15ex\hbox {/}}
\newcommand{\bvdir}{\bar{v}\kern -5.75pt\raise 0.15ex\hbox {/}}
\newcommand{\Ddir}{D\kern -7.75pt\raise 0.20ex\hbox {/}}
\newcommand{\ldir}{l\kern -5.0pt\raise 0.2ex\hbox{/}}
\newcommand{\varepsdir}{\varepsilon\kern -5.5pt\raise 0.15ex\hbox{/}}

\def\str{\operatorname{str}}
\def\tr{\operatorname{tr}}
\def\Tr{\operatorname{Tr}}

\def\diag{\operatorname{diag}}
\def\negcdot{\negmedspace\cdot\negmedspace}

\begin{document}

\preprint{IJS/TP-6/03; LPT Orsay 03-19}

\title{$B\to \pi$ and $B\to K$ transitions in partially  quenched chiral perturbation theory}

\author{Damir Be\'cirevi\'c}
 \email{Damir.Becirevic@th.u-psud.fr}
\affiliation{%
Laboratoire de Physique Th\'eorique (B\^at 210), Universit\'e Paris Sud,  
Centre d'Orsay, 91405 Orsay-Cedex, France}

\author{Sa\v{s}a Prelov\v{s}ek} 
\email{Sasa.Prelovsek@ijs.si}
\affiliation{
J.~Stefan Institute, Jamova 39, P.O. Box 3000, 1001 Ljubljana, Slovenia\\
and Department of Physics, University of Ljubljana, Jadranska 19, 1000 Ljubljana, Slovenia
}

\author{Jure Zupan}
 \email{zupan@physics.technion.ac.il}
\affiliation{%
Department of Physics,  Technion--Israel Institute of Technology, Technion City, 32000 Haifa, Israel}

\date{19 May 2003}

\begin{abstract}
We study the properties of the $B\to \pi$ and $B\to K$ transition form factors in partially quenched QCD 
by using the approach of partially quenched chiral perturbation theory
combined with the static heavy quark 
limit. We show that the form factors change almost linearly when varying the value of the sea quark mass, 
whereas the dependence on the valence quark mass  contains both
the standard and chirally divergent (quenched) logarithms.  A simple strategy 
for the chiral extrapolations in the lattice studies with $\ns=2$ is suggested. 
It consists of the linear extrapolations from the realistically 
accessible quark masses, first in the sea and then in the valence
quark mass. From the present approach, we estimate the uncertainty induced by 
such extrapolations to be within $5\%
$. 
\end{abstract}

\maketitle

\section{\label{Introduction}Introduction}

In order to extract the Cabibbo-Kobayashi-Maskawa matrix element $\vert V_{ub}\vert $ from the ex\-pe\-ri\-men\-ta\-lly measured 
decay rate for $B\to \pi \ell \nu_\ell$ at Belle and {\sc BaBar}, a reliable theoretical prediction of 
the corresponding hadronic matrix element is indispensable.  
Two major sources of systematic uncertainty in the current lattice QCD calculations of the 
$heavy\to light$ pseudoscalar meson transitions are the quenched approximation and the errors 
associated with the chiral extrapolations.  All the available results of the lattice studies of the semileptonic 
$B\to \pi$ decay are obtained in the quenched approximation ($N=0$) and 
by working with light quark masses larger than $m_s^{phys}/2$~\cite{lattice}.

Since fully unquenched QCD lattice simulations  (i.e., with $N=3$) are not feasible it is important to have a 
method  to assess whether  or not a given physical quantity is prone to large quenching errors. With such an
ambition in mind,  Sharpe~\cite{sharpe},  and later Bernard and Golterman~\cite{bernard}, formulated the quenched chiral 
perturbation theory (QChPT). 
By  confronting the predictions derived in QChPT  with those obtained in the full chiral perturbation theory (ChPT), 
one gets a rough estimate on the size of quenching errors. The approach has been extended to the heavy-light quark systems by 
combining ChPT and the heavy quark effective theory (HQET)~\cite{booth,zhang}. The effect of complete 
quenching on the lattice determination of the $B\to \pi$ form factors has been studied recently by the 
present authors in Ref.~\onlinecite{JSD}. 
It has been found that the quenching errors on the $B\to \pi$ and $B\to K$ transition form factors may be 
uncomfortably large (typically larger than $20\%
$). This conclusion somewhat spoils  the significance of the
impressive agreement that has been reached amongst various lattice 
groups using  di\-ffer\-ent lattice techniques to compute $B\to \pi$
decay in the quenched approximation~\cite{lattice}.

It is therefore highly important to perform the simulations in which the effects of dynamical quarks are included.
A step in that direction is to implement the partial 
quenching, i.e., to include in the simulation
$N_{\rm sea}$ dynamical light quarks with masses generally different
from those of the valence quarks. 
The lattice studies with $N_{\rm sea}=2$ are likely to be performed first, which is the main motivation for the present 
work. 

We calculate the form factors for $B\to \pi$ and $B\to K$ transitions in partially quenched chiral perturbation theory 
(PQChPT)~\cite{Bernard:1993sv,Damgaard:1998xy,Sharpe:2001fh}, combined with HQET~\cite{zhang}. 
We work in the static heavy quark limit and at next-to-leading order
(NLO) of the chiral expansion. This approach is valid for small recoil momenta  
($v\negcdot p$), i.e., the same one currently accessible from the lattice simulations. 
We limit ourselves to the case of $\ns$ degenerate sea quarks of mass $\ms$. On the basis of our calculation we conclude that the dependence of the 
form factors on the sea quark mass is essentially linear, and that the form factors are finite 
as $\ms \to 0$ when $\mv$ is non-zero. On the other hand, the limit $\mv\to 0$, 
with $\ms \ne \mv$, is not well defined, since in this case the form
factors contain the chirally divergent 
``quenched" logarithmic terms $\ms \ln \mv$. Our analysis of PQChPT with 
$\ns=2$ shows that a simple linear chiral extrapolation first
in the sea quark mass, $\ms$, and then in the valence quark mass, $\mv$, introduces
an  extrapolation error of only  $5 \%
$, where the linear extrapolations are made from the range of the quark 
masses that are currently accessible in the lattice simulations.  
Furthermore, by assuming that the low energy constants in the full ChPT with $N=2$ and $N=3$ are equal,  we deduce the residual quenching errors of lattice studies with 
$\ns=2$ to be in the ball park of $10-20\%
$. Those two conclusions indicate that, in comparison with the fully quenched simulation, 
the additional computational cost of making the partial unquenching (with $\ns =2$) is well 
worth the effort.

The remainder of this paper is organized as follows. In Sec.~\ref{pQChPT} we remind the reader about 
the elements of the PQChPT and set our notation. In Sec.~\ref{sec:chi-results} we present the results 
of the NLO calculation in PQChPT for $B\to \pi(K)$ form factors. 
The quenching errors and chiral extrapolation formulas are discussed in
Sec.~\ref{chiral-extrap}. Main conclusions are shortly stated in 
Sec.~\ref{conclusions}.

\section{\label{pQChPT}Partially quenched chiral perturbation theory}
\setcounter{equation}{0}

In this section we make a succinct summary of  PQChPT. We 
consider partially quenched QCD with $N_{\rm sea}$ sea 
 quarks degenerate in mass ($\ms$), and with two valence quarks $q_{a}$ and $q_{b}$ of 
 masses $m_a$ and $m_b$, respectively. Partial quenching is introduced 
by adding bosonic ``ghost"-quarks $\tilde{q}_{a,b}$ of spin $\tfrac{1}{2}$ and mass $m_{a,b}$, which 
cancel the fermion loops of {\it valence} quarks~\cite{sharpe,bernard,Bernard:1993sv,morel}. 
Assuming the  spontaneous symmetry breaking pattern to be the same as
in full QCD~\cite{Sharpe:2001fh}, 
the leading-order Lagrangian for the (pseudo) Goldstone 
bosons is~\cite{Bernard:1993sv,Damgaard:1998xy,Sharpe:2001fh}
\bea \label{light}
{\cal L}_{\rm light}&=&{f^2\over 8}\str \left(\partial_\mu \Sigma \partial^\mu
\Sigma^\dagger \right) + {f^2 \mu_0\over 2}
\str({\cal M}\Sigma+{\cal M}\Sigma^\dagger)\cr
&&+{\cal L}_4\,,
\eea
with $f \approx 130$~MeV, $\str(U)={\displaystyle{\sum_{i}}} U_{ii}\epsilon_i$, and 
\bea
\epsilon_i&=&\diag(\underbrace{1,\dots,1}_{N_{\rm sea}+2},-1,-1),\cr 
{\cal M}&=&\diag(m_a,m_b,\underbrace{\ms,\dots,\ms}_{N_{\rm sea}},m_a,m_b) , 
\eea
while
\bea \label{eqQ:2} 
\Sigma = \exp\left(2i\frac{\Phi}{f}\right), 
\qquad \Phi=
\begin{pmatrix}
\phi&\chi^\dagger\\
\chi&\tilde \phi\;
\end{pmatrix}\;.
\eea
The graded ${\rm SU}(2\!+\!N_{\rm sea}|2)$ matrix $\Phi$ contains a  
$[(2+N_{\rm sea})\times(2+N_{\rm sea})]$ matrix $\phi$ of quark--anti-quark 
bosons, a $[2\times 2]$ matrix $\tilde\phi$ of 
ghost--anti-ghost bosons, and the matrices of pseudoscalar fermions, 
$\chi^\dagger\sim \bar{\tilde{q}} q$ and  
$\chi\sim \bar q \tilde q$.  In this framework the heavy $\eta^\prime$ has been integrated out 
($m_0\to\infty$) in a way similar to the full ChPT. This is to be contrasted to the fully 
quenched theory, where the $\eta^\prime$ cannot 
be integrated out. The propagators of the ``off-diagonal" and ``diagonal"  mesons ($\Phi_{ab}\sim \bar q_a q_b$) are given respectively 
by
\bea
&&\langle\Phi_{ab}|\Phi_{ba}\rangle_{a\not = b}=i \frac{ \epsilon_a}{p^2-M_{ab}^2},\cr
&&\langle\Phi_{aa}|\Phi_{bb}\rangle =i \biggl[\frac{\delta_{ab} \epsilon_a}{p^2-M_{aa}^2}-\frac{1}{N_{\rm sea}} 
\frac{p^2-M_{S}^2}{(p^2-M_{aa}^2)(p^2-M_{bb}^2)}\biggr], \cr
&& \eea
with $M_{ab}^2=2\mu_0 (m_a+m_b)$, and $M_{S}^2=4\mu_0 \ms$.

Since PQChPT with $N_{\rm sea}=3$,  and $m_a=m_b\equiv \ms$ coincides with the full ChPT with 
$m_u=m_d=m_s$, the corresponding low energy constants obtained in the
two theories are equal.~\footnote{Note, however, that in the usual definition
of low energy constants~\cite{gasser} an ${\rm SU}(3)$ relation between
operators is
used~\cite{Donoghue:dd}, which is not valid for the general
${\rm SU}(N_1|N_2)$ case. This subtlety does not concern us here, as it does not
change the definition of low energy constants entering in our
calculations.}
Since the low energy constants do not depend 
on the quark masses, the mentioned equality among low energy constants  
persists even when the valence and sea quark masses of partially quenched theory are not of 
the same size~\cite{Sharpe:2001fh}. On the other hand the low energy constants do depend 
on the number of dynamical quark flavors and therefore the equality holds 
only for $N_{\rm sea}= 3$.

The two terms appearing at NLO in the effective Lagrangian~\eqref{light} relevant 
to $B\to \pi$ and $B\to K$ form factors are
\bea \label{eqQ:1}
{\cal L}_4 &=& 4  \mu_0 \Big\{ L_4 \str\left(\partial_\mu \Sigma \partial^\mu 
\Sigma^\dagger\right)\str\left({\cal M}\Sigma^\dagger+\Sigma 
{\cal M}^\dagger\right) \cr
&&+ L_5 \str\left[\partial_\mu \Sigma^\dagger \partial^\mu 
\Sigma\left({\cal M}\Sigma^\dagger+\Sigma {\cal M}^\dagger\right)\right]+\cdots \Big\}~.\;
\eea
It is straightforward to verify that Eq.~(\ref{light}) leads to  
the standard, full QCD,  chiral Lagrangian after setting  
$\str\to \tr$, $\Phi\to\phi$ (for a review  
of ChPT see Ref.~\onlinecite{Donoghue:dd} or any paper listed in Ref.~\onlinecite{chiral-reviews}).

\subsection{Incorporating the heavy quarks}
To extend the applicability of the Lagrangian~\eqref{light} to the heavy-light mesons,   
the heavy quark spin symmetry needs to be included. This is achieved by combining the pseudoscalar 
($P^a$) and vector ($P^{\ast\ a}_\mu$) 
heavy-light mesons in one field: 
\bea
H_a(v) &=& {1 +  \vdir \over 2} \left[ P^{\ast\ a}_\mu (v)\gamma_\mu - P^a (v)\gamma_5\right]~, \cr
 \overline H_a(v) &=& \gamma_0 H_a^\dagger (v) \gamma_0\;.
\eea 
We also introduce the covariant derivative and the axial field as 
\bea
D_\mu^{ba}H_b &=&  \partial_\mu H_a -  H_b {1\over 2}[ \xi^\dagger \partial_\mu \xi + 
\xi \partial_\mu \xi^\dagger ]_{ba}~,\cr 
{\bf A}_\mu^{ab}
&=& {i\over 2}[ \xi^\dagger \partial_\mu \xi - 
\xi \partial_\mu \xi^\dagger ]_{ab} \;,
\eea
where $a$ and $b$ run over the light quark flavors, and $\xi = \sqrt{\Sigma}$. 
Finally, the partially quenched chiral 
Lagrangian for the heavy-light mesons in the static heavy quark limit reads~\cite{booth,zhang} 
\bea \label{heavy}
{\cal L}_{\rm heavy}&=&-\str_a \Tr[\overline{H}_a i v \negcdot D_{ba}H_b]\cr
&& \hfill \cr
&&+g 
\str_a\Tr[\overline{H}_aH_b \gamma_\mu 
{\bf A}_{ba}^\mu \gamma_5]
+{\cal L}_3 \;,
\eea
where $g$  is the coupling of the heavy meson doublet to 
the Goldstone boson. 
The higher order terms in the expansion in $v\negcdot p$  and in $m_q$ [${\cal O}(p^2)$] 
have the following form~\cite{booth,JSD}:
\bea \label{eqQ:3}
{\cal L}_3 &=&2\lambda_1\str_a \Tr[\overline H_aH_b]
({\cal M_+})_{ba}\cr
&& \hfill \cr
&&+k_1\str_a \Tr[\overline{H}_a i v \negcdot D_{bc}H_b]
({\cal M_+})_{ca}\cr
&& \hfill \cr
&&+ k_2\str_a \Tr[\overline{H}_a i v \negcdot D_{ba}H_b]\str_c({\cal M_+})_{cc}+\dots\;,
\eea
with ${\cal M_+}={1\over 2}(\xi^\dagger M\xi^\dagger+\xi M\xi)$. We again display only the terms 
that contribute to the heavy-to-light form factors of $B\to \pi,K$ transitions. In the above equations, 
``Tr" stands for the trace over Dirac indices, whereas ``str" is the super-trace over the light flavor 
indices.  The standard chiral Lagrangian for heavy-light mesons~\cite{reviews-HQChPT,burdman} is recovered by  replacing 
$\str\to \tr$, $\Phi\to\phi$.

The bosonized {\it heavy} $\to$ {\it light} weak current $(V-A)$, in the static heavy quark 
limit and at NLO in the chiral expansion, reads~\cite{booth,JSD}
\bea
\begin{split}
\label{current}
 J^\mu\equiv \bar{q}_a\gamma^\mu(1-\gamma_5)Q &\to \frac{i \alpha}{2}\Tr[\gamma^\mu (1-\gamma_5) H_b] \xi_{ba}^\dagger  \\
&\hspace*{-12mm}+\frac{i \alpha}{2}\varkappa_1\Tr[\gamma^\mu (1-\gamma_5) H_c ]\xi_{ba}^\dagger ({\cal M_+})_{cb}\\
&\hspace*{-12mm}+\frac{i 
\alpha}{2}\varkappa_2\Tr[\gamma^\mu (1-\gamma_5) H_b] \xi_{ba}^\dagger \str_c({\cal M_+})_{cc}\;.
\end{split}
\eea

 The phase of the heavy meson can be chosen 
in such a way that  the constants $\alpha$,
$\varkappa_1$ and $\varkappa_2$ are real. At the leading order in the chiral expansion, 
the constant $\alpha$ is proportional to the heavy-light meson decay constant, $\alpha=\sqrt{m_B} f_B$.

\subsection{Form factors}

The standard decomposition of the vector current matrix element between two pseudoscalar meson states is
\bea\label{par1}
\langle P (p)\vert \bar q\gamma_\mu b \vert B(p_B)\rangle &=& 
\left[ (p_B + p)_\mu - q_\mu {m_B^2 -
m_P^2\over q^2} \right]  F_+(q^2) \cr
&&
+ {m_B^2 - m_P^2 \over q^2} q_\mu
F_0(q^2)\;,
\eea
where the form factors $F_{+,0}$ are functions of the momentum transfer squared $q^2=(p_B-p)^2$. 
A light meson $P$ stands for $\pi$, $K$ with the light quark in the current being $q=d,s$, respectively. 

We work in the static limit ($m_B\to \infty$) in which    
the eigenstates of QCD and HQET Lagrangians are  related through
${1\over \sqrt{m_B}} \vert B(p_B)\rangle_{\rm QCD} =  \vert B(v)\rangle_{\rm HQET}$.
In the static limit it is more convenient to use a definition in which the form factors are 
independent of the heavy meson mass. We will use
\bea \label{par2}
\langle P (p)\vert \bar q \gamma_\mu b_v \vert B(v)\rangle_{\rm{ HQET}} &=& 
\bigl[ p_\mu - (v\negcdot p) v_\mu\bigr] f_p (v\negcdot p)\cr 
&&+\ v_\mu  f_v (v\negcdot
p)\;,
\eea
where the field $b_v$ does not depend on the heavy quark mass. 
The form factors $f_{p,v}$ are functions of the variable 
\bea
v\negcdot p = { m_B^2 + m_P^2 - q^2 \over 2 m_B}\;,
\eea
which is the energy of the light meson $E_P$, in the heavy meson rest frame. 
The relation between the quantities defined in Eqs.~(\ref{par1}) and (\ref{par2}) 
is obtained by  matching QCD to HQET at the scale $\mu\sim m_b$~\cite{hill}.  
By setting the matching constants to their tree level values and neglecting the subleading 
terms in the heavy quark expansion, one has
\bea
\label{f}
&&\Biggl.  F_0(q^2) \Biggr|_{q^2 \approx
q^2_{\rm max}} =  {1\over \sqrt{m_B}}  f_v(v\negcdot p)\;,\cr 
&& \hfill \cr 
&& \hfill \cr 
&&\Biggl. F_+(q^2) \Biggr|_{q^2 \approx
q^2_{\rm max}}={\sqrt{m_B} \over 2} f_p(v\negcdot p)\,,
\eea
i.e., the usual heavy mass scaling 
laws for the semileptonic form factors~\cite{isgur}.~\footnote{ Notice that in the heavy 
quark limit the tensor current form factor,  $F_T(q^2)$, defined as 
$\langle P(p) \vert \bar q\sigma_{\mu \nu}q^\nu b \vert B(p_B)\rangle =
i  \left[ q^2 (p_B + p)_\mu - (m_B^2
- m_P^2) q_\mu \right]F_T(q^2)/(m_B + m_P)$, 
is related to the vector form factor via the Isgur-Wise relation~\cite{isgur},  $F_T(q^2)= F_+(q^2)$.}

\section{\label{sec:chi-results}Expressions for the form factors in PQChPT}

\begin{figure}
\begin{center}
\epsfig{file=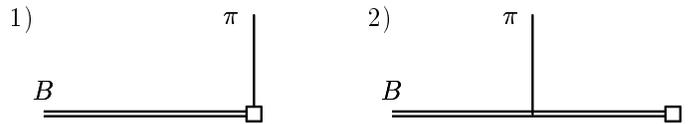, height=1.65cm}
\caption{\label{fig1}  The point 1) and the pole 2) 
tree level Feynman diagrams contributing to 
$\text{\it heavy} \to \text{\it light}$ 
transition form factors. The box denotes the weak current insertion.} 
\end{center}
\end{figure}
\begin{figure*}[th!!]
\begin{center}
\epsfig{file=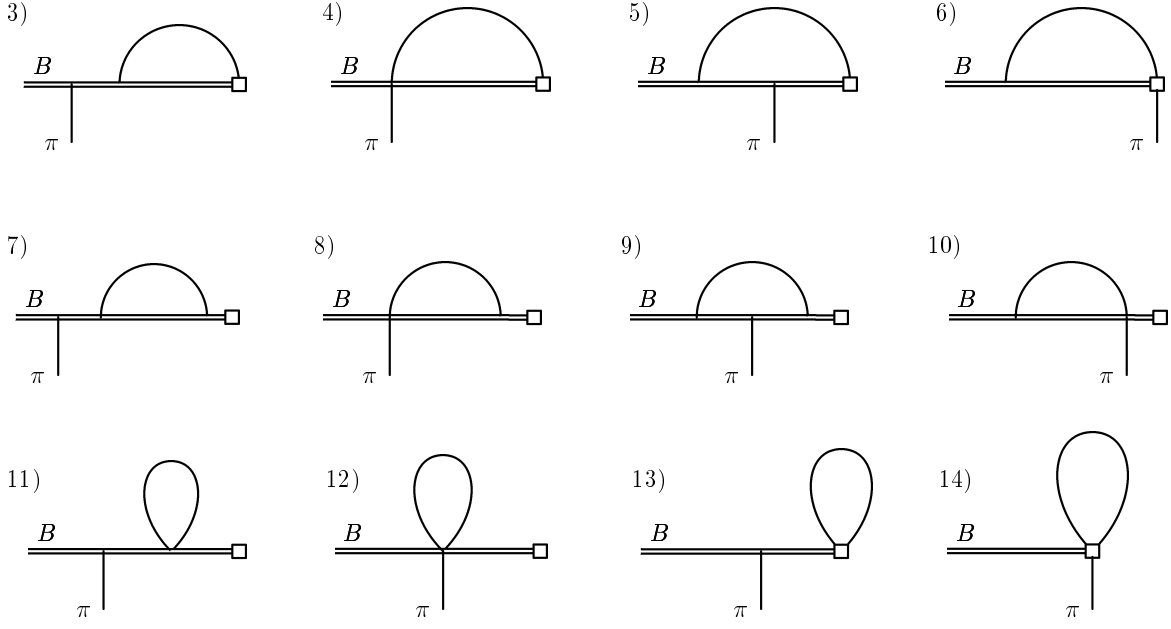, height=8.9cm}
\caption{\label{fig2} The one loop  contributions to 
the $B\to \pi $ transition. Double/single lines denote the heavy/light meson, 
while the weak current insertion is depicted by the empty box.} 
\end{center}
\end{figure*}
In this section we give the expressions for the form factors $f_p(v\negcdot p)$ and 
$f_v(v\negcdot p)$ as derived in PQChPT with  $N_{\rm sea}$ degenerate quarks 
of mass $\ms$, and with the valence light quarks of masses $m_a$ and 
$m_b$. When necessary, we will use   $P_{ba}$ to denote the light 
pseudoscalar meson with the valence quark content 
$ P_{ba}\sim q_b\bar q_a$ and  $B_a\sim b\bar q_a$ for the heavy mesons.

The tree level expressions for $B_a\to P_{ba}$ transition form factors 
are given by the point and pole diagrams in Fig.~\ref{fig1}, which give rise to 
 $f_v$ and $f_p$, respectively
\bea
f_p^{\text{Tree}} (v\negcdot p) = {\alpha \over f} {g \over v\negcdot p + \Delta_b^\ast} \;, 
\qquad f_v^{\text{Tree}} (v\negcdot p) = {\alpha \over f} \;, 
\eea
where  $\Delta_b^\ast =m_{B_b^*}-m_{B_a}$.
Although the heavy quark spin symmetry suggests $\Delta_b^\ast \to 0$, we 
 keep $\Delta_b^\ast $ finite in the tree-level form factors because it provides the pole to 
the form factor $f_p$  at $m_{B_b^*}^2$.~\footnote{The pole dominance 
is easily seen if one rewrites the denominator of $f_p (v\negcdot p)$ as 
$v\negcdot p + \Delta^\ast = ({m_{B^\ast}/ 2}) \left( 1 - q^2/m_{B^\ast}^2 \right)$, 
where the corrections $(m_{B^*}-m_B)/m_B$ and $m_P^2/m_B^2$ are neglected.}

The NLO chiral corrections to the form factors are conveniently expressed as
\begin{align}\label{fpvNLO}
f_{p,v}^{B_a\to P_{ba}} = f_{p,v}^{\text{Tree}}[ 1&+(\delta f_{p,v})^{\text{Loop}} \nonumber\\ &+
c_a^{p,v} m_a+c_b^{p,v} m_b+c_{\rm sea}^{p,v} \ms ]\,,
\end{align}
where
\begin{subequations}
\begin{align}
c_a^p&=c_a^v-\varkappa_1=\frac{1}{2}k_1-\frac{16\mu_0 L_5}{f^2}\  ,\\
c_b^p&=c_b^v+\varkappa_1=\varkappa_1-\frac{16\mu_0 L_5}{f^2}\ ,\\
c_{\rm sea}^p&=c_{\rm sea}^v=\biggr(\varkappa_2+\frac{1}{2}k_2 -\frac{32\mu_0 L_4}{f^2}\biggr)N_{\rm sea}\ ,
\end{align}
\end{subequations}
that arise from the ${\cal O}(p^3)$~\eqref{eqQ:1} and ${\cal O}(p^4)$~\eqref{eqQ:3} terms in the Lagrangian, 
as well as from the weak 
current~\eqref{current}. The loop contributions are written as
\beq \label{deltafpv}
(\delta f_{p,v})^{\text{Loop}}=\sum_{I} \delta f_{p,v}^{(I)}+\frac{1}{2} \delta Z_B^{\text{Loop}}+
\frac{1}{2} \delta Z_P^{\text{Loop}}\,,
\eeq
where the sum runs over all the graphs depicted in Fig.~\ref{fig2}, and the last two terms arise from the loop 
contributions to the wave-function renormalizations. The explicit expressions  for the loop corrections 
in Eq.~\eqref{deltafpv} are rather lengthy, and we relegate them to Appendix~\ref{app:B}. In the calculation of the 
loop integrals we used the naive dimensional regularization, and the $\msbar +1$ 
renormalization prescription, 
i.e. we subtract $\bar \Delta= 2/(4-d)- \gamma+\ln(4\pi)+1$~\cite{gasser}.  We neglect  mass 
differences $\Delta$ between $B_a$, $B_b$, $B_a^\ast$ and $B_b^\ast$ meson states whenever they 
appear in loops. We also remark that no dependence of the form factors $f_v$ and $f_p$ on $v\negcdot p$ arises 
from the counterterms, so the  modification of the tree level $v\negcdot p$ dependence is entirely due to the 
chiral loop corrections.

Finally, we explicitly checked that the expressions for the form factors obtained in PQChPT with 
$N_{\rm sea}=3$ and $m_a=m_b=\ms$, indeed agree with the ones obtained in full ChPT with $m_u=m_d=m_s$ 
(for the full ChPT expressions see Ref.~\onlinecite{JSD}).

\subsection*{\label{chir_lim}Form factors in the chiral limit}

In this section we make several important remarks concerning the
chiral behavior of the form factors at some fixed (albeit small) $v\negcdot p$. We focus on the situation in which 
the light pseudoscalar meson consists of degenerate  valence quarks, i.e.  $M_{V}^2=4\mu_0 \mv$.
We distinguish the following three cases.
\begin{itemize}
\item[(1)]  Expansion of $(\delta f_{p,v})^{\text{Loop}}$ \eqref{fpvNLO} for $\ms\to 0$ 
and fixed nonzero $\mv$, results in a linear term in $\ms$, but without logarithmic terms, i.e.
\begin{equation}
 \label{ms_dep}
\delta f_{p,v}^{\text{Loop}}(M_{S}^2)=C_0^{p,v}+ C_2^{p,v} M_{S}^2+\dots,
\end{equation}
where $M_{S}^2=4\mu_0 \ms$. The coefficients $C_0^{p,v},C_2^{p,v}$ are functions
of $v\negcdot p$ and $M_{V}$, with ellipses representing higher terms in
the expansion. This result is helpful for the future lattice
simulations with $N_{\rm sea}\neq 0$. Namely, it suggests that 
a linear extrapolation  of the form factors in $\ms$  from the directly accessible sea quark masses 
down to $\ms \to 0.5\times (m_u +m_d)^{phys.}$ is not modified by the chiral logarithms, 
provided the valence quark mass is kept fixed. In the lattice studies, the
constants $C_0$ and $C_2$ are then to be obtained by fitting the data to 
Eq.~\eqref{ms_dep}. 
We will scrutinize this point more in the next section.

\item[(2)] If, on the other hand, one keeps the sea quark mass
$\ms$ nonzero  and studies the 
limit  $\mv\to 0$, the chiral logarithms appear. In addition to the
ordinary chiral logarithms, i.e., of the form  $M_{V}^2 \ln(M_{V}^2)$, one also picks the quenched logarithmic divergences 
$M_{S}^2 \ln(M_{V}^2)$. In this limit 
 \begin{align}
 \label{mv_dep}
 \delta f_p^{\text{Loop}}=\frac{1}{(4 \pi f)^2 \ns}\biggl[&\biggl(-2 g^2 \frac{M_{S}^2}{(v\negcdot p)^2}+ 
 1+3 g^2\biggr) M_{V}^2 \ln(M_{V}^2) \nonumber\\
&\hspace*{-5mm}-\frac{1+3 g^2}{2} M_{S}^2 \ln(M_{V}^2) -4 \pi g^2 \frac{M_{S}^2}{v\negcdot p} M_{V} \biggr]\nonumber\\
&+ {C'}_0^{p}+{C'}_2^{p}M_{V}^2+\dots\;,\nonumber\\
&\hfill  \\
\delta  f_v^{\text{Loop}}=\frac{1}{(4 \pi f)^2 \ns} &\frac{1+ 3 g^2}{2} (2 M_{V}^2-M_{S}^2) \ln(M_{V}^2)
\nonumber\\
&+{C'}_0^{v}+{C'}_2^{v}M_{V}^2+\dots\;,\nonumber
\end{align} 
where the coefficients ${C'}_{0,2}^{p,v}$ are functions of $v\negcdot p$ and $M_{S}$. We remind the reader that 
$M_{V}^2=4\mu_0 \mv$.

\item[(3)] The case with  $\mv=\ms$ is actually the case of full,
unquenched,  QCD. The form factors in the chiral limit, $\mv=\ms\to 0$,
are finite and coincide with the ones derived in the standard (unquenched) ChPT with $N_{\rm sea}$
mass-degenerate quarks. In this limit the chiral behavior is
 \begin{align}
\label{ChPtexp}
 \delta f_{v,p}^{\text{Loop}}=\frac{1}{(4 \pi f)^2}\biggl[&\big(1+3
g^2 \big)\frac{(1-\ns^2)}{2\ns} M_{V}^2\ln\left( M_{V}^2\right)\nonumber \\
&+{C''}_0^{v,p}+{C''}_2^{v,p}M_{V}^2+\dots\biggr]\;.
 \end{align}
\end{itemize} 
Notice that the situation for the $B\to
K$ transition is qualitatively very similar to the $B\to \pi$ case, described above.

\section{Results (``phenomenology")\label{chiral-extrap}}

As we mentioned in the introduction, the current quenched lattice simulations of the $B\to \pi$ matrix element are 
performed with the light quark masses $m_q \gtrsim m_s^{\text{phys}}/2$, where $m_s^{\text{phys}}$ 
is the physical strange quark mass. This feature is likely to remain as such in the forthcoming (partially) unquenched 
lattice studies. Moreover, the first partially quenched lattice simulations will most probably be performed 
with $\ns=2$. For that reason, in what follows, we consider the case
of two degenerate sea quarks ($\ns=2$).

Once the partially quenched lattice QCD results become available, the 
NLO chiral expressions for $B\to \pi$ form factors (provided in
Sec.~\ref{sec:chi-results}) can be used to extrapolate from the light 
quark masses directly 
accessed in the lattice simulations down to the physical $u/d$-quark mass.

One way to proceed is to match the chiral expressions~\eqref{fpvNLO} with 
the lattice data at some intermediate values of $(m_{\rm val.})_M$ and 
$(m_{\rm sea})_M$, at which the lattice data are used to fix the low 
energy constants (cf. discussion in Sec.~6.2 of Ref.~\onlinecite{JSD}). 
From that point down to the chiral limit, the extrapolation is made by 
using such determined constants, plus the coefficients of the chiral 
logarithms predicted in PQChPT. The matching procedure is needed to 
make contact of the  pronounced linear light quark mass dependence 
of the (quenched) lattice data with the NLO expressions derived in 
ChPT, thus guiding the extrapolation to the physical pion mass.~\footnote{The linear dependence of the 
$B\to\pi$ form factors in the light quark mass is observed for the quenched data as well as for the 
$B\to 0$ transition matrix element in both quenched and partially quenched studies. }
It is, however, not clear at which point the above-mentioned matching should be made, i.e., 
that for the masses lighter than that used in the matching procedure one can trust the chiral 
perturbation theory. Is it $m_s^{\text{phys}}$,
$m_s^{\text{phys}}/2$, or $m_s^{\text{phys}}/4$? Clearly, depending
on what we choose for $(m_{\rm val.})_M$ and $(m_{\rm sea})_M$, from which
we include the logarithmic terms in the extrapolation, we will get different results for the
physically interesting form factors. Moreover, since the lattice data show linear 
dependence on quark masses this also means that the variation of matching point 
will introduce a larger variation of the extrapolated values, if the
ChPT dependence is very nonlinear~\cite{JSD,Becirevic:2002mh}. If instead  
the behavior of the form factor predicted by ChPT is close to 
linear, the precise point where we match ChPT expressions to the lattice data 
will not matter at all, as long as both the size
and the slope (the leading and the NLO coupling of ChPT expression) are
matched to the lattice data.
It is thus far more reasonable to look for the strategy to extrapolate in $m_{\rm val.}$ and
$m_{\rm sea}$, in which the ChPT expressions exhibit almost linear
dependence. This is precisely where the discussion made in Sec.~\ref{sec:chi-results} ``Form factors 
in the chiral limit" becomes important. Recall that we found that the dependence of $B\to \pi$ form
factors on $m_{\rm sea}$ (with other variables fixed) is linear, whereas the one on 
 $m_{\rm val.}$ does include nonlinear terms. However, one can suppress
the most dangerous chirally divergent term $m_{\rm sea}\ln m_{\rm
val.}$ if the $m_{\rm sea}$ is close to the chiral limit. 
This suggests that a fairly linear behaviour can be obtained if one first extrapolates in 
$m_{\rm sea}$, and then in $m_{\rm val.}$. It is this observation that we will
elaborate more on in the present section.

In the numerical evaluations we shall assume that the low energy constants appearing in PQChPT 
with $N_{\rm sea}=2$ are equal to their counterparts in the full ChPT with 
$N=3$. In other words, we assume that the low energy constants 
depend weakly on the number of sea quarks, $\ns$. In addition, and for an easier comparison, 
we will take the same values for the low energy constants as in Ref.~\onlinecite{JSD}:
$\alpha=0.56(4)~\gev^{3/2}$, $g=0.5(1)$, $f=0.13\ \gev $, $\mu_0=1.14(20)\ \gev $, $L_4= -(0.5 \pm 0.5) \times 10^{-3}$, and  
$L_5= (0.8 \pm 0.5)\times 10^{-3}$. The counterterms $L_4$ and $L_5$ are evaluated at the renormalization scale $\mu=1$~GeV, 
which is also the choice for $\mu$ in the loop integrals. For more details about the choice of parameters 
and the complete list of references we refer the reader to  Ref.~\onlinecite{JSD}. 
For the other counterterms $\varkappa_{1,2}$, $k_{1,2}$, there are no
available experimental data or any indications from the 
lattice data and we will take them to be zero. We have verified that the impact of the variation of 
these constants  on our conclusions is insignificant: 
(i) the loop corrections depend only on $g$ and are not sensitive to variations of $g$ in the range $g=0.5(1)$,
(ii) the counterterms have negligible  effect on the $B\to \pi$ form factors, while they can modify the magnitude
of $B\to K$ form factors but not their dependence on $v\negcdot p$ [Eq.~\eqref{fpvNLO}].    
For the physical strange quark mass we take $m_s^\msbar(2\ \gev)\approx 90$ MeV~\cite{wittig}.

\subsection{$B\to \pi$ transition}

We work in the isospin limit and set the masses of the pion valence quarks to be equal, 
$m_a=m_b\equiv \mv$. 
\begin{figure}
\begin{center}
\epsfig{file=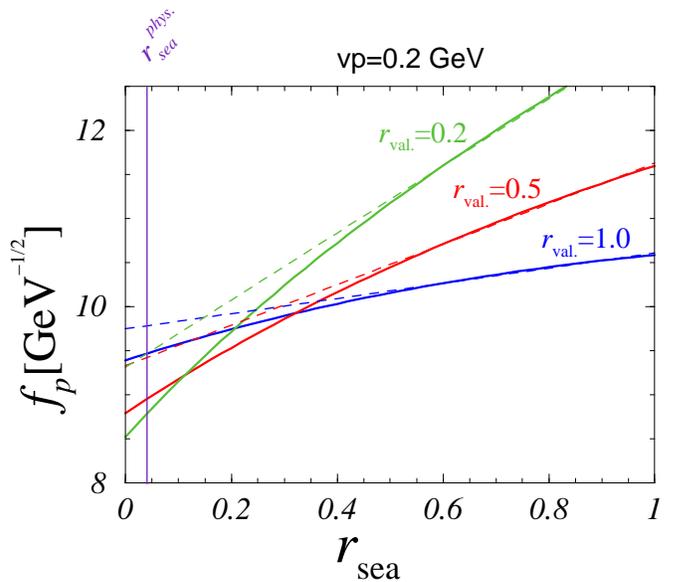,width=8.6cm}
\caption{ \label{figS}  Comparison of complete
$f_{p}^{B\to \pi}$ expressions in the partially quenched $\ns=2$ theory 
and the linear extrapolations. Here $v\negcdot p=0.2$ GeV and $\rv=0.2,0.5, 1.0$ 
are held fixed, while $\rs$ is varied. Linear fit is made to the PQChPT expressions 
from Appendix~B in the range $\rs\in [0.5,1.0]$, and then extrapolated to the chiral limit. 
The vertical line marks the physical value $\rs=1/25$. }
\end{center}
\end{figure}
To examine the chiral behavior of the form factors we will use 
\beq
\rv=\mv/m_s^{\text{phys}}, \qquad \rs=\ms/m_s^{\text{phys}}~,
\eeq
dimensionless parameters defined with respect to the physical strange quark mass. 
In the lattice studies, lowering the sea quark mass, $\ms$ (i.e., $\rs$), is numerically costlier than
lowering the valence quark mass, $\mv$  (i.e., $\rv$). Fortunately, as discussed in
the previous section, the dependence of $f_{p,v}(v\negcdot p)$ on $\rs$ is expected to
be linear, so that even larger extrapolations in $\ms$ may still lead to
reasonably small extrapolation errors. This is illustrated in 
Fig.~\ref{figS}  for the case of the form factor $f_p$, with $v\negcdot p=0.2$~GeV. 
We see that for small $\rs < \rv$ and $\rs \to 0$, the form factor 
is indeed a linear function of $\rs$, as already stated in Eq.~\eqref{ms_dep}. That 
behavior gets modified when $\rs > \rv$, resulting in the smooth forms that are very close 
to linear. If we linearly extrapolate the behavior of the form factors from the range of sea quark mass 
$\rs \in [0.5, 1]$ down to the chiral limit, we observe only a small discrepancy with respect to 
the complete expression for $f_{p}$ (of about $5$~\%). 
The same observation holds for the form factor $f_v$, with the discrepancy between the extrapolated and 
the result of the complete expression being  approximately halved  with respect to the one estimated for  
$f_p$. As it can be observed from Fig.~\ref{figS}, the uncertainty induced by the extrapolation in 
$\ms$ gets larger as the valence quark mass is decreased. However, for the quark masses that are 
realistic in the current lattice studies, $\rv \gtrsim 0.5$, the mentioned uncertainty is below the $5~\%
$ level.
\begin{figure}
\begin{center}
\epsfig{file=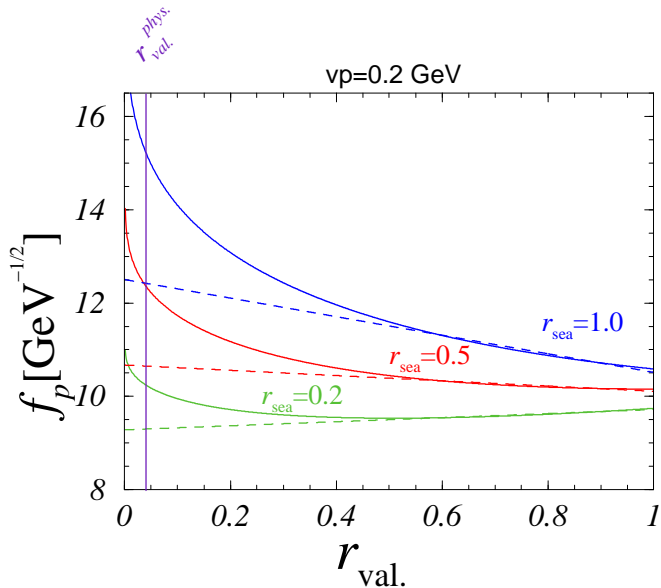,width=8.6cm}
\caption{ \label{figV} The situation similar to what is plotted in Fig.~{\rm\ref{figS}}, but with 
$\rs$ and $\rv$ exchanging the roles, i.e., $\rv$ is varied for three fixed values of $\rs=0.2,0.5, 1.0$.}
\end{center}
\end{figure}

The other possibility, i.e., to extrapolate $f_{p,v}$  in $\rv$ for a fixed value of $\rs$, is 
illustrated in Fig.~\ref{figV}. We see that in this case the difference between the linear 
extrapolation and the complete expression is much more pronounced, which reflects the  
presence of the quenched chiral logarithm $\propto \rs\ln(\rv)$, and hence the effect is larger for 
the heavier sea quark masses (cf. Eq.~\eqref{mv_dep}). 
Similar observations to the ones shown in Figs.~\ref{figS} and \ref{figV}, are also valid for the form factor $f_v$.

\begin{figure}
\begin{center}
\epsfig{file=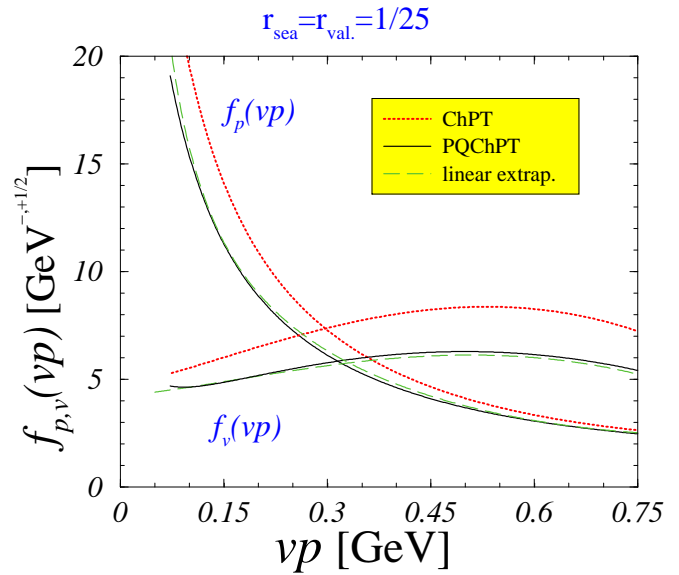,width=8.6cm}
\caption{ \label{compare} Solid lines show  
$f_{p,v}^{B\to \pi}(v\negcdot p)$ in the partially quenched theory
with $\ns=2$ at physical point $\rs=\rv=1/25$. The dashed lines are obtained by
first linearly extrapolating in $\rs$ then in $\rv$ to the physical point, as explained in the text. 
For an easier comparison we also
plot the same form factors obtained in ChPT with $N=3$ by a dotted line 
(using the same values
for low energy constants as in PQChPT with $\ns=2$). }
\end{center}
\end{figure}

An interesting observation is that the dependence of $f_{p,v}$ on
$\rv$ (with $\rs$ and $v\negcdot p$ fixed) is such that the form factors 
at the physical point $\rv\simeq 1/25$ are {\it larger} than what one gets
from  the linear extrapolation. 
On the other hand, when $\rv$ and $v\negcdot p$ 
are fixed, the form factors at the physical point $\rs\simeq 1/25$ are {\it smaller} than 
the linearly extrapolated ones.

Therefore by first performing a linear
extrapolation in $\rs$, and then in $\rv$ will not only allow one to avoid  
the spurious quenched logarithms, but it will also produce a mutual cancellation 
of the  errors induced by the two chiral extrapolations.  
Indeed, the errors of the two consecutive 
extrapolations, as shown on Fig.~\ref{compare}, are strikingly small. 
Performing linear fits
to the  $\ns=2$  PQChPT expressions for $\rs, \rv \in [0.5,1]$ leads
to the errors below $5\%
$, for both $f_p(v\negcdot p)$ and $f_v(v\negcdot p)$ form factors, 
and for a large range of $v\negcdot p \in [0.05, 0.75]$~GeV. If, however, the chiral extrapolation is made linearly by keeping the valence 
and sea quark masses equal, $\rv=\rs$, the resulting error is in the $20\%
$ range due to the explicit chiral logarithmic corrections given in Eq.~\eqref{ChPtexp}.

 For the reader's convenience, in Fig.~\ref{compare}, we also plot 
the result of the fully unquenched theory with $N=3$. We observe a pleasant feature that the 
shapes of the form factors in these two theories are very similar. 
Notice also that the results obtained in PQChPT with $\ns=2$ are systematically lower 
from the ones obtained with ChPT $N=3$ by  $10\%
-20$\%. It should be stressed, 
however, that this discrepancy might be artificial and merely a consequence of our assumption that the 
low energy constants do not change significantly when $N=2\to 3$. 
More information on the low energy constants in the theory with $N=2$ is
needed to get a clearer picture on the significance of the differences depicted in Fig.~\ref{compare}.

\begin{figure}
\begin{center}
\hspace*{-7mm}\epsfig{file=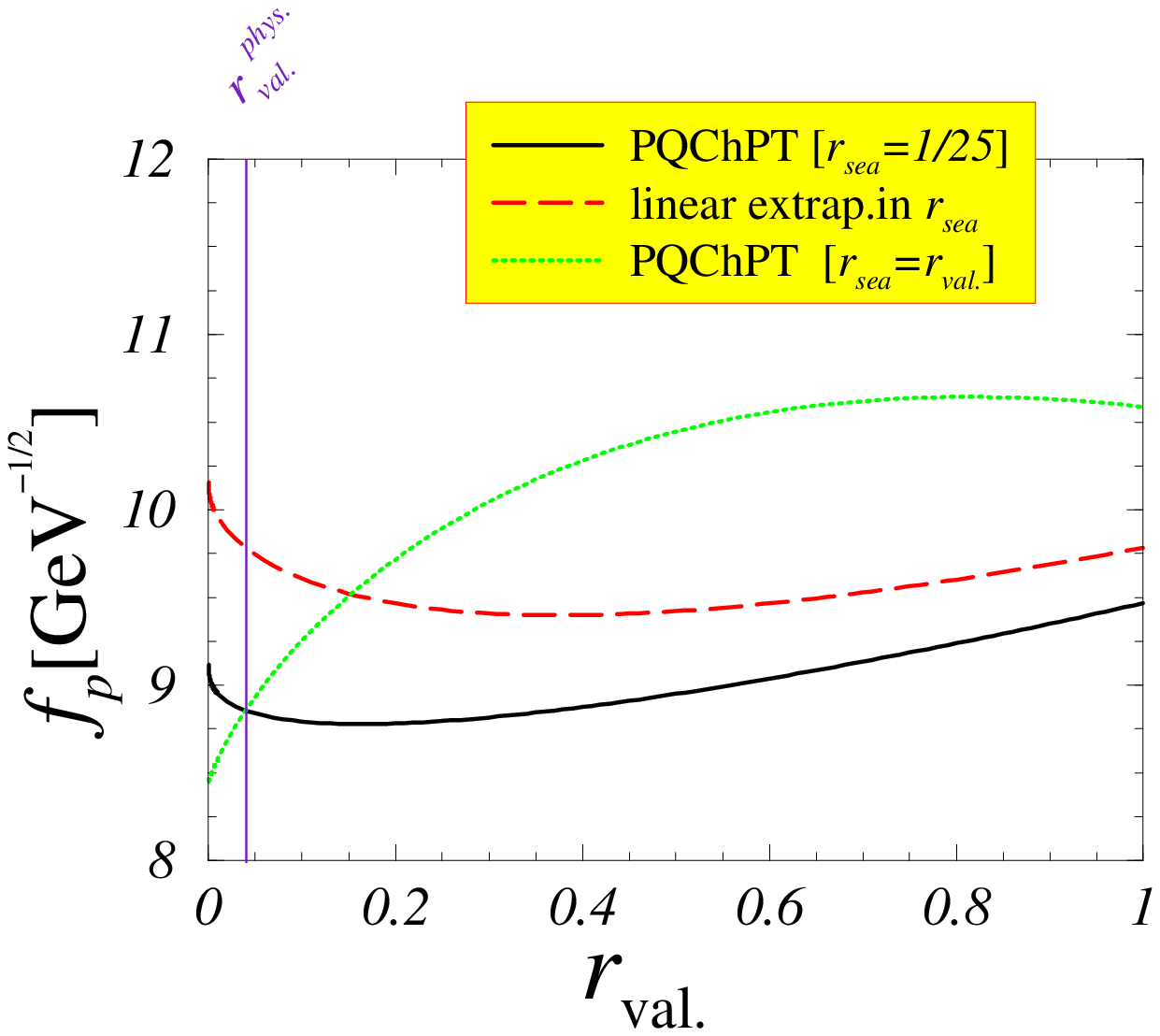,width=8.3cm}
\epsfig{file=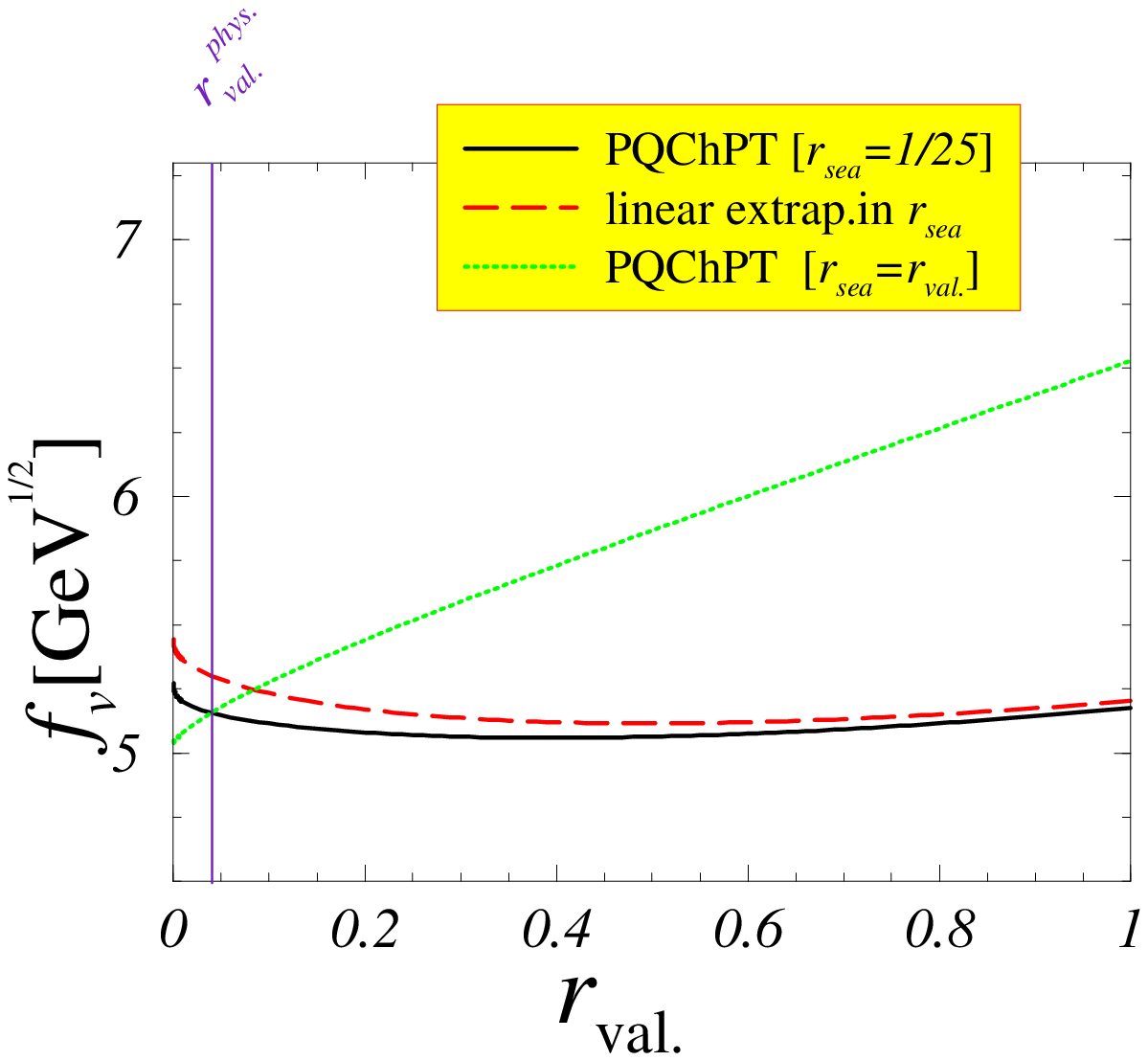,width=8.05cm}
\caption{ \label{fp_r.fig} 
Solid lines show the $\rv$  dependence of $f_{p,v}^{B\to \pi}$ [Eq.~(\ref{f})] 
in the partially
quenched theory with $\ns=2$ and at $\rs=1/25$, $v\negcdot p=0.2\ \gev$. 
The dashed lines show the form factors  linearly extrapolated from
$\rs\in[0.5,1.0]$ down to physical point $\rs=1/25$. The dotted lines
refer to the dependence on $\rv$ while keeping $\rs=\rv$. }    
\end{center}
\end{figure}

Another lesson comes from Fig.~\ref{fp_r.fig}, where we plot the $\rv$
dependence of the form factors  by using the complete $\ns=2$ expressions, derived in PQChPT, 
and by linearly extrapolating in $\rs$ as discussed above. 
We see that the error due to the use of linear extrapolation of $f_p$ in $\rs$ 
 is fairly small at large $\rv$,
but then increases as $\rv$ approaches its physical value. Reducing
$\rv$ below $\rv < 0.5$ in future lattice studies will therefore improve
on the simple consecutive linear extrapolations (first in $\rs$, then
in $\rv$, for $\rs,\rv\in[0,5,1.0]$) \underline{\it only if}, together with
smaller $\rv$, also the smaller values of $\rs$  are reached in the simulation. 
For the high precision studies with both $\rs$ and $\rv$ well below $0.5$, it might even
be more advantageous to use extrapolations with $\rs=\rv$ (cf. Fig.~\ref{fp_r.fig},
left), contrary to what we advocated for more realistic studies (at present), namely with $\rs,\rv\gtrsim 0.5$.

Finally, we consider the soft pion theorem, which states that the
ratio $F_0/(f_B/f_\pi) \to 1$ as $v\negcdot p$ and quark masses go to
zero in full ChPT. We verified that this is satisfied also in
partially quenched theory in the limit in which $v\negcdot p\to 0$, and $\ms=\mv\to
0$. Numerically, the ratio stays within $0.95<[F_0/(f_B/f_\pi)]<1.25$ in
the ranges of $v \negcdot p <0.2~\gev$ and $0<\rv,\rs< 1$, but it
diverges for  $\mv \to 0$ if $\ms \neq \mv$.

\subsection{ Ratios $f_{p,v}^{B\to K}/f_{p,v}^{B\to \pi}$ }

\begin{figure}
\begin{center}
\hspace*{-7mm}\epsfig{file=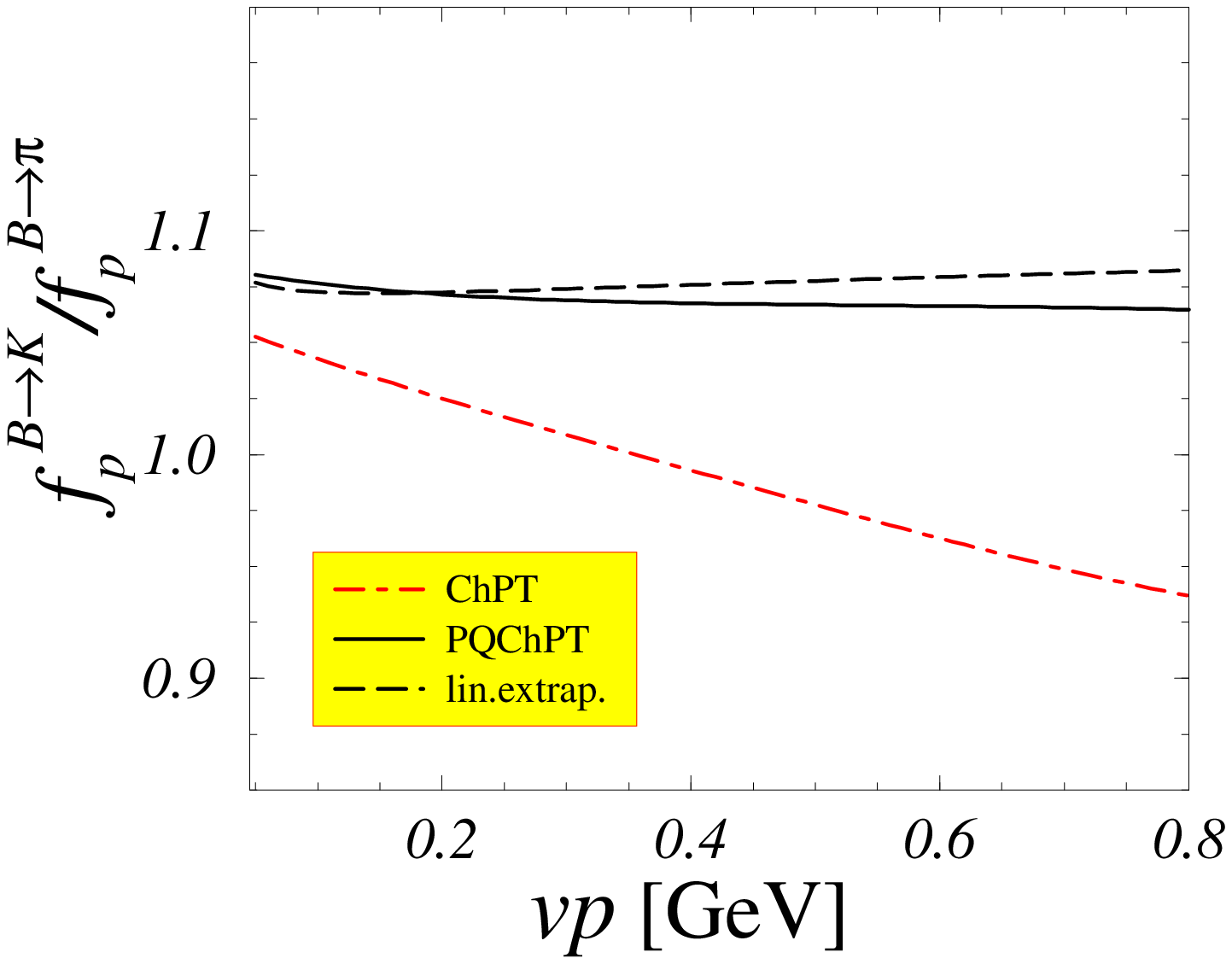,width=8.2cm}
\epsfig{file=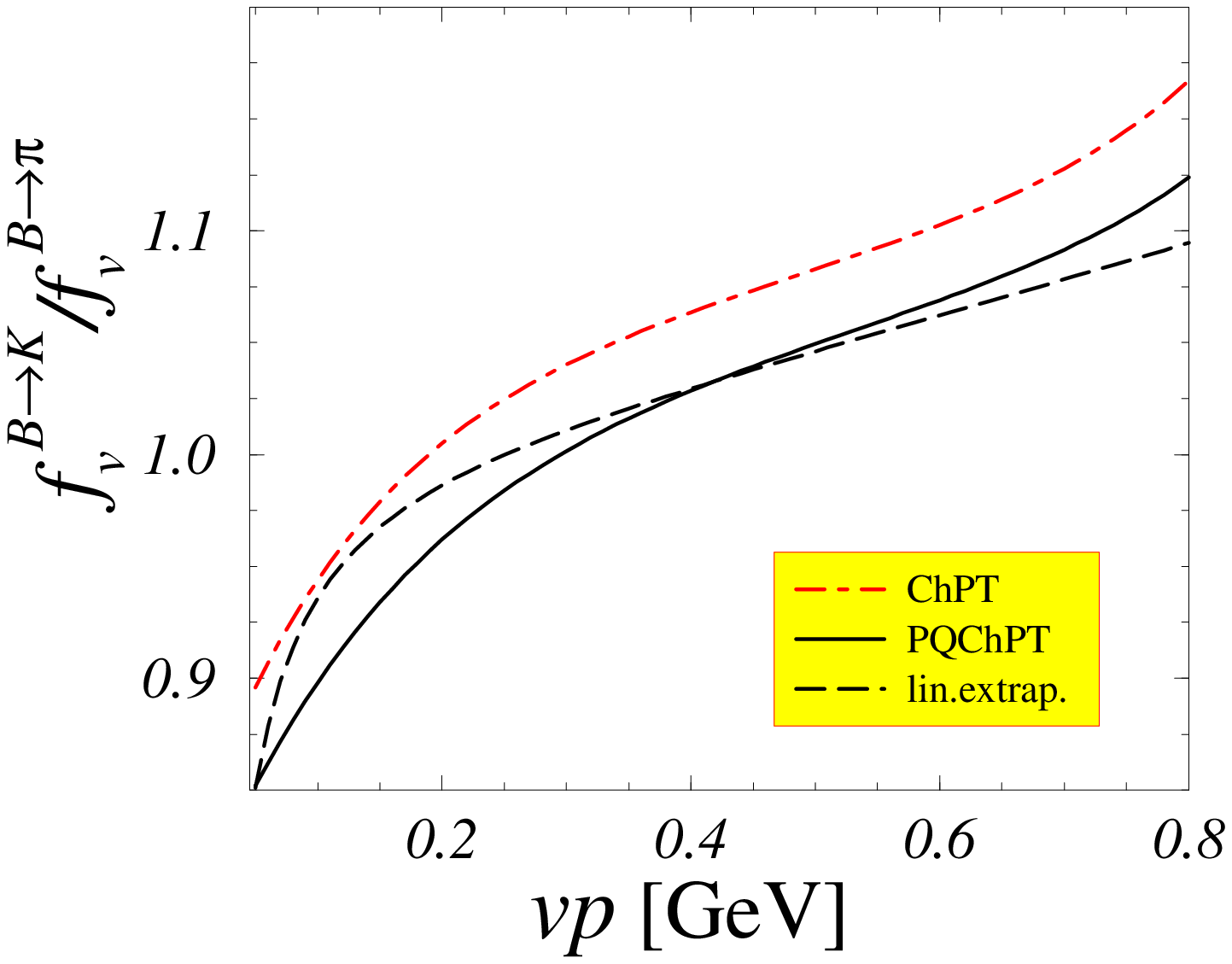,width=8.2cm}
\caption{ \label{raz_fp_r.fig}  Behavior of the SU(3) breaking ratio 
$R_{p,v}=f_{p,v}^{B\to K}/f_{p,v}^{B\to \pi}$ (at physical values of
quark masses) as functions of $v\cdot
p$.  The prediction of PQChPT with 
$\ns=2$ is denoted by the solid lines. The dashed curves are obtained after 
consecutive extrapolations, first in $\rs$ and then in $\rv$ (as explained in the text), 
while the dot-dashed curve is the prediction of the full ChPT with $N=3$. }
\end{center}
\end{figure}

Instead of repeating the previous discussion for the $B\to K$ form factors, we investigate the ratios 
$R_{p,v}=f_{p,v}^{B\to K}/f_{p,v}^{B\to \pi}$. Our kaon is composed of valence quarks with one mass 
fixed at the physical strange quark mass and the other varied,
i.e., $M_K^2= 2\mu_0 m_s^{\text{phys}} (1+\rv) $, whereas 
the pion valence content reads, $M_\pi^2= 4\mu_0 m_s^{\text{phys}} \rv $. 
We also ignore the difference between the poles in $f^{B\to K}_p$ and in 
$f^{B\to \pi}_p$, i.e., we take $m_{B^\ast}=m_{B_s^\ast}$.

We again observe the linear dependence of the ratio $R_{p,v}$ under 
varying the light sea quark mass, $\ms$, while keeping fixed values of $\mv$ and 
$v\negcdot p$.  As in the previous section we see that 
the variation of $\mv $ in the chiral limit  at fixed  $\ms$ exhibits both 
standard and quenched chiral logarithmic dependences. Therefore, like in the 
$B\to \pi$ case, a consecutive linear extrapolation in $\rs$ and then in $\rv$ 
leads to very small errors of extrapolation, namely below 5\% 
(see Fig.~\ref{raz_fp_r.fig}).

In Fig.~\ref{raz_fp_r.fig} we observe that the  dependence of $R_p$ on 
$v\negcdot p$, as inferred from the PQChPT with  $\ns=2$, 
differs from the one obtained in the 
full ChPT with $N=3$ at the $10\%
$ level. Note that this feature does not change under 
the variation of ${\cal O}(p^3)$, ${\cal O}(p^4)$
 low energy constants  given in Eqs.~\eqref{eqQ:3} and \eqref{eqQ:1}. In addition, the
$v\negcdot p$ dependence of $R_{p}$ in partially quenched theory does
not change significantly when varying $\rs$ in the range
$\rs\in (0,1)$. It is important to stress that the apparent disagreement between 
$R_{p}^{\rm PQChPT}$ and  $R_p^{\rm ChPT}$ as functions of $v\negcdot p$  
disappears after replacing $\ns=2$ by $\ns =3$ in our PQChPT expressions.        
Finally, it should be mentioned that the absolute values of $R_{p,v}$ (but not their $v\negcdot p$ behavior) 
may be further affected by the uncertainties in $\varkappa_1$ and $L_5$, because these constants enter 
the expressions for $R_{p,v}$, multiplied by $m_K^2$.

\section{\label{conclusions}Conclusions}

In this paper we presented the expressions for the $heavy \to light$ pseudoscalar meson form factors
as derived at NLO in PQChPT and in the static limit of HQET. This calculation is useful in the perspective of 
the QCD simulations on the lattice, since the most wanted $B\to \pi$ transition form factors are 
expected to be computed quite soon in the partially quenched QCD with $\ns=2$. The approach adopted in this paper is
valid for small recoils (i.e., close to $q^2\to q^2_{\rm max}$), the part of the phase space that is 
also accessible by the current lattice simulations. 
The main benefit of the expressions derived in PQChPT lies in the fact that one can disentangle 
the chiral behavior in the sea quark mass, $\ms$, from the chiral dependence in the valence light quark 
mass, $\mv$.  Our results clearly suggest a linear behavior of the form factors under the
variation of the sea quark mass value $\ms$, when   
the valence quark mass $\mv$ is kept fixed. On the other hand, if the
sea quark mass $\ms$ is fixed to some value that is 
directly accessible from the current lattice studies with $\ns=2$ (e.g. $\rs \gtrsim 0.5$), the form factors 
exhibit both the standard and  quenched chiral logarithms when approaching the physical point, $\rv \simeq 1/25$. 
For that reason,  we propose a strategy of computing the $B\to \pi
(K)$ semileptonic form factors  on the lattice (in partially quenched theory with
$\ns=2$ degenerate light quarks)  
that requires a computation to be performed at several values 
of the sea quark mass with the valence quark mass held fixed. 
The extrapolation of $F_{+,0}(q^2)$ in the sea quark mass to the
physical $u$/$d$ quarks can then be made linearly at fixed valence quark mass. 
This should then be followed by an extrapolation in the valence quark mass. In this way the quenched logarithm is  
avoided since the pathological term is of the form $\ms \ln(\mv)$. As a result,  the above 
procedure leads to results which are different by no more than $5\%
$ with respect to the complete prediction of PQChPT  with $\ns=2$. 
This makes a strong call for the partially quenched lattice computations of the $heavy \to light$ 
transition matrix elements with $\ns=2$.

Regarding the comparison of the form factors obtained in PQChPT with $\ns=2$ and the ones 
obtained in ChPT with $N=3$, 
we notice that the results obtained with $N=2$ are systematically lower than their $N=3$ counterparts. 
The difference is in the ball park of $10\%
-20\%
$. 
At this point it is not clear whether this difference is a realistic
estimate of the present approach, or merely a consequence of our lack of knowledge of the values of the 
low energy constant in the theory with $N=2$. The above estimate of $10\%
-20$\% 
is obtained by assuming the low energy constants to be independent of the number of flavors.


\begin{acknowledgments}
We thank V.~Lubicz  for comments on the manuscript of the present paper.
The work of S.P. and J.Z. has been supported in part by the Ministry 
of Education, Science and Sport of the Republic of Slovenia. 
D.B. acknowledges a partial support of the E.C.'s contract HPRN-CT-2000-00145 
``Hadron Phenomenology from Lattice QCD". Laboratoire de Physique Th\'eorique is 
unit\'e mixte de Recherche du CNRS - UMR 8627.
\end{acknowledgments}

\appendix

\section{\label{appA}Chiral loop integrals}
In this appendix we list the dimensionally regularized integrals encountered 
in the course of calculation. For more details, see Ref.~\onlinecite{jure} and references therein. 
\begin{align} \label{int-A}
i\mu^\epsilon\int \frac{d^{4-\epsilon}q}{(2\pi)^{4-\epsilon}
}\frac{1}{q^2-m^2}&=\frac{1}{16 \pi^2}I_1(m)\;, \nonumber \\
i\mu^\epsilon\int \frac{d^{4-\epsilon}q}{(2\pi)^{4-\epsilon}}
\frac{1}{(q^2-m^2)(q\negcdot v-\Delta)}&=\frac{1}{16
\pi^2}\frac{1}{\Delta}I_2(m,\Delta)\;,
\end{align}
where
\begin{align}
I_1(m)&=m^2 \ln\Bigl(\frac{m^2}{\mu^2}\Bigr)-m^2\bar{\Delta}\; , \nonumber \\
I_2(m,\Delta)&=-2\Delta^2 \ln\Bigl(\frac{m^2}{\mu^2}\Bigr)-4 \Delta^2
F\Bigl(\frac{m}{\Delta}\Bigr) +2 \Delta^2(1+\bar{\Delta})\; ,
\end{align}
where $\bar{\Delta}=2/\epsilon -\gamma +\ln(4\pi)+1$. The function $F(x)$ has 
been calculated in Ref.~\onlinecite{stewart}, for both the negative and positive values of
the argument:
\begin{equation}
F\left(\frac{1}{x}\right)= \left\{
\begin{aligned}
-\frac{\sqrt{1-x^2}}{x}&\left[\frac{\pi}{2}-\tan^{-1}\left(\frac{x}{\sqrt{1-x^2}}\right)\right], \,
&|x|\le 1\\
\frac{\sqrt{x^2-1}}{x}&\ln \left(x+\sqrt{x^2-1}\right),\, &|x|\ge 1
\; .
\end{aligned}\right.
\end{equation}
In addition to the integrals~(\ref{int-A}), one also needs the following two:
\bea
&&i\mu^\epsilon\int \frac{d^{4-\epsilon}q}{(2\pi)^{4-\epsilon}}
\frac{q^\mu}{(q^2-m^2)(q\negcdot v-\Delta)}\nonumber \\
&&=\frac{v^\mu}{16
\pi^2} \left[ I_2(m,\Delta)+I_1(m) \right]\;, \\
&& i\mu^\epsilon\int \frac{d^{4-\epsilon}q}{(2\pi)^{4-\epsilon}}
\frac{q^\mu q^\nu}{(q^2-m^2)(q\negcdot v-\Delta)}\nonumber \\
&&=\frac{1}{16
\pi^2}\Delta\left[ J_1(m,\Delta)\eta^{\mu \nu}+J_2(m,\Delta)v^\mu
v^\nu \right] \;,
\eea
with
\begin{subequations}\label{eq-1}
\begin{align}
\begin{split}
J_1(m,\Delta)=&(-m^2+\frac{2}{3}\Delta^2)\ln\left(\frac{m^2}{\mu^2}\right)+\frac{4}{3}(\Delta^2-m^2)
F\left(\frac{m}{\Delta}\right)\\
&-
\frac{2}{3}\Delta^2(1+\bar{\Delta})+\frac{1}{3}m^2(2+3\bar{\Delta})+\frac{2}{3}m^2-\frac{4}{9}\Delta^2
\; ,
\end{split}
\\
\begin{split}
J_2(m,\Delta)=&(2
m^2-\frac{8}{3}\Delta^2)\ln\left(\frac{m^2}{\mu^2}\right)-\frac{4}{3}(4
\Delta^2-m^2)F\left(\frac{m}{\Delta}\right)\\
&+\frac{8}{3}\Delta^2(1+\bar{\Delta})-\frac{2}{3}m^2(1+3\bar{\Delta})-\frac{2}{3}m^2+
\frac{4}{9}\Delta^2
\; .
\end{split}
\end{align}
\end{subequations}
The functions $J_1(m,\Delta),J_2(m,\Delta)$ differ from the ones in
Ref.~\onlinecite{boyd} by the last two terms in
Eq.~\eqref{eq-1} which are of ${\cal O}(m^2, \Delta^2)$. These
additional (finite) terms originate from the fact that $\eta^{\mu \nu}$ is
the ($4-\epsilon$)-dimensional metric tensor.

\section{\label{app:B}Loop contributions to form factors in PQChPT}
In this appendix we list expressions for the one-loop chiral corrections to the form factors $f_{p,v}$ 
[Eq.~\eqref{deltafpv}]  
in partially quenched ChPT (the form factors for fully quenched ChPT and full ChPT are collected in Appendix B of Ref.~\onlinecite{JSD}).  
The  formulas apply to  the $B_a\to P_{ba}$ transition with $B_a\sim b\bar q_a$ and $P_{ba}\sim q_b\bar q_a$.  
They are expressed in terms of the  pseudoscalar meson masses
$M_{a,b,S}^2 = 4 \mu_0 m_{a,b,{\rm sea}}$, $M_{ab}^2 = 2 \mu_0 (m_a+m_b)$, and $M_{aS,bS}^2 = 2 \mu_0 (m_{a,b}+\ms)$.  
We list the nonzero contributions to $f_{p,v}^{(I)}$, with the superscript ``$I$" corresponding to a label of 
the diagram shown in Fig.~\ref{fig2}. 
\begin{widetext}
\bea 
\delta f_p^{(7)} &=& \frac{3 g^2}{(4 \pi f)^2}\    
\Big\{ \ns \left[  J_1(M_{bS}, v\negcdot p)-\frac{1}{v\negcdot p} \frac{2\pi}{3} M_{bS}^3\right]  -\frac{1}{\ns}
 \left( 1+ \left(
M_b^2-M_S^2\right)\frac{\partial}{\partial M_b^2}\right) 
\left[ J_1(M_b, v\negcdot p)-\frac{1}{v\negcdot p} \frac{2\pi}{3} M_b^3\right] \;,\nonumber \\
&& \hfill \nonumber \\
\delta f_p^{(9)}&=& -\frac{ g^2}{(4 \pi f)^2}\frac{1}{\ns}\frac{1}{(M_a^2-M_b^2)}
\bigg[ \left(M_a^2-M_S^2\right)\left( J_1(M_a,v\negcdot p) -\frac{1}{v\negcdot p}\frac{2 \pi}{3}M_a^3\right) -\left(M_b^2-M_S^2\right)\left(J_1(M_b,v\negcdot p)-\frac{1}{v\negcdot p}\frac{2 \pi}{3}M_b^3\right)
\bigg]\;,\nonumber \\
 \hfill \nonumber \\
 \delta f_p^{(12)} &=&- \frac{1}{6 (4 \pi f)^2} \biggl\{ 
\ns\Big(I_1(M_{aS})+I_1(M_{bS})\Big) -\frac{1}{\ns}\Big(G_1(M_a,M_S)+G_1(M_b,M_S)\Big)\biggr. 
+\frac{2}{\ns}H_1(M_a,M_b,M_S)\biggr\} \;, \nonumber \\
&& \hfill \nonumber \\
\delta f_p^{(13)} &=& -\frac{1}{2 (4 \pi f)^2}
\Big[\ns I_1(M_{bS})-\frac{1}{\ns}G_1(M_b,M_S) \Big]\,,\hspace*{-12mm}\label{fpQ}
\eea
and
\bea
\delta f_v^{(4)}&=&\frac{1}{2 (4 \pi f)^2} \biggl\{ 
 \ns\Big[I_1(M_{bS})+2 I_2(M_{bS},v\negcdot p)\Big]+\frac{1}{\ns}\Big[H_1(M_a,M_b,M_S)- G_1(M_b,M_S)\Big] 
 \nonumber
\\
 && \hspace*{6.35cm}+\frac{2}{\ns}\Big[H_2(M_a,M_b,M_S,v\negcdot p)-G_2(M_b,M_S,v\negcdot p)\Big] \biggr\}\,, \nonumber
\\
&& \hfill \nonumber \\
\delta f_v^{(14)}&=&-\frac{1}{6 (4 \pi f)^2}\biggl\{ 
\ns\Big[I_1(M_{aS})+I_1(M_{bS}) \Big]-\frac{1}{\ns}\Big[G_1(M_a,M_S)+G_1(M_b,M_S)\Big] -\frac{1}{\ns} H_1(M_a,M_b,M_S) \biggr\}~. \nonumber
\\ &&
\eea
The one loop chiral  corrections to the wave function renormalization factors $Z_{B,P}$ are
\bea
\delta Z_{P_{ab}}^{\text{Loop}}&=&
\frac{1}{3 (4 \pi f)^2}\biggr\{\ns\Big[ I_1(M_{aS})+I_1(M_{bS})\Big]+\frac{2}{\ns}H_1(M_a,M_b,M_S) - 
\frac{1}{\ns}\Big[G_1(M_a,M_S)+G_1(M_b,M_S)\Big]\biggr\}\,, \nonumber \\
\delta Z_{B_a}^{\text{Loop}}&=&\frac{3 g^2}{(4 \pi f)^2}\biggr\{-\ns I_1(M_{aS})+\frac{1}{\ns} G_1(M_a,M_S)\biggr\}~.
\eea
\end{widetext}

The functions $I_1(m)$,  $I_2(m)$, and $J_1(m,\Delta)$ are given in Appendix A. In addition, we have used the abbreviations
\begin{align}
G_1(M,M_S)&=\frac{\partial}{\partial M^2}\left[(M^2-M_S^2)I_1(M)\right]\,,\\
H_1(M_a,M_b,M_S)&=\frac{1}{M_a^2-M_b^2}\left[\left(M_a^2-M_S^2\right)I_1(M_a)\right.\nonumber \\
& -\left.\left(M_b^2-M_S^2\right)I_1(M_b)\right]\,,
\end{align}
and similarly for $G_2(M,M_S,v\negcdot p)$ and $H_2(M_a,M_b,M_S,v \negcdot p)$, that are obtained by simply replacing $I_1(M) \to 
I_2(M,v\negcdot p)$ in the above expressions.  Note that in the limit of degenerate valence quarks, $M_a\to M_b$, we have 
\begin{align}
\lim_{M_a\to M_b}H_1(M_a,M_b,M_S)&=G_1(M_a,M_S)\,,\\
\lim_{M_a\to M_b}H_2(M_a,M_b,M_S,v \negcdot p)&=G_2(M_a,M_S,v\negcdot p)\,.
\end{align}

We reiterate that  
the mass differences ($\Delta$) among $B$, $B^\ast$, $B_s$, and $B_s^\ast$ mesons have been consistently neglected in the loops, 
since  $v\negcdot p > \Delta^{\ast}$, for the $B\to \pi, K$ 
transitions.
This induces a spurious singularity in the expression for the diagrams (7), at $v\negcdot p \to 0$. To handle such singularities we followed 
the proposal by Falk and Grinstein~\cite{falk} and resum the corresponding diagrams and then simply subtract the term that would renormalize 
the $B^\ast$-meson mass. 

\newpage 


\end{document}